\documentclass[lettersize,journal]{IEEEtran}
\usepackage{amsmath,amsfonts}
\usepackage{algorithmic}
\usepackage{array}
\usepackage[caption=false,font=normalsize,labelfont=sf,textfont=sf]{subfig}
\usepackage{textcomp}
\usepackage{stfloats}
\usepackage{url}
\usepackage{xcolor}
\usepackage{verbatim}
\usepackage[mathscr]{eucal}
\usepackage{graphicx}
\def\BibTeX{{\rm B\kern-.05em{\sc i\kern-.025em b}\kern-.08em
    T\kern-.1667em\lower.7ex\hbox{E}\kern-.125emX}}
\usepackage{balance}

\usepackage{amsthm}

\newtheorem{lemma}{Lemma}
\newtheorem{definition}{Definition}

\begin{document}
\title{Send Message to the Future? Blockchain-based Time Machines for Decentralized Reveal of Locked Information}
\author{
    \IEEEauthorblockN{Zhuolun Li\IEEEauthorrefmark{1}, Srijoni Majumdar\IEEEauthorrefmark{1}, Evangelos Pournaras\IEEEauthorrefmark{1}}
    \\\IEEEauthorblockA{\IEEEauthorrefmark{1}School of Computing, University of Leeds}
    \thanks{Manuscript created October 2024; This work was developed by the School of Computing, University of Leeds. The opinions expressed here are entirely those of the authors. No warranty is expressed or implied. User assumes all risk.}
}

\maketitle

\begin{abstract}
Conditional information reveal systems automate the release of information upon meeting specific predefined conditions, such as a designated time in the future. By designing a new practical timed-release cryptography architecture and a secret sharing scheme with verifiable information reveal, a novel data sharing system is devised on smart contracts that ``sends messages in the future" with highly accurate decryption times. Using the same architecture, this paper also introduces a breakthrough in the understanding, design, and application of conditional information reveal systems that are highly secure and decentralized. A complete evaluation portfolio is provided to this pioneering paradigm, including analytical results, a validation of its robustness in the Tamarin Prover and a performance evaluation of a real-world, open-source system prototype deployed across the globe. Using real-world election data, we also demonstrate the applicability of this innovative system in e-voting, illustrating its capacity to secure and ensure fair electronic voting processes. 
\end{abstract}


\section{Introduction}
\IEEEPARstart{C}{onditional} information reveal uses computer systems to automatically reveal information upon the verification of certain requirements. From releasing information after a certain point of time~\cite{timothy_timed-release_1993, rivest_time-lock_1996}, to disseminating information based on geographical location~\cite{proof_of_witness_presence, location_verification_application_1, location_verification_application_2} and restricting access to classified documents based on identity~\cite{boneh2001identity, signle_sign_on}, these are all instances of information reveal dictated by a certain condition of time, location, or access right. One such condition, pivotal in its role across a multitude of applications, is time.

The study of time-based conditional information reveal systems is known as timed-release cryptography or time lock encryption. It guarantees the confidentiality of a message until a specific point in time, upon which the message is automatically decrypted and becomes accessible to the intended parties. Despite its practical potential, its journey from theory to widespread usage has been hindered by the need to ensure accurate decryption time, message confidentiality against adversarial parties, and energy efficiency~\cite{timothy_timed-release_1993, rivest_time-lock_1996}.

This paper introduces a smart-contract-based timed-release cryptography system with improvements in security and efficiency compared to the existing proposals~\cite{celebi2018kimono, ning2018keeping, li2018decentralized, li2021silentdelivery, bacis2021told, ID-based_TRE}. Moreover, this paper shows how general conditional information reveal systems can be secured using the same architecture as the proposed timed-release cryptography system.

Smart contract and threshold cryptography are used in the proposed system to allow a consortium of distributed secret holders to safeguard sensitive information with a minimum level of trust. The decentralization, immutability, and transparency nature of blockchains ensures authenticity, availability and assurance in the communication among clients and secret holders. Furthermore, a blockchain-based incentive mechanism is applicable to ensure secret holders' honesty, thereby enhancing the overall system reliability. 

Notably, this paper proposes a secret sharing scheme with reveal-verifiability for conditional information reveal systems to enhance their efficiency and security. The common definition of verifiable secret sharing (VSS) and publicly verifiable secret sharing (PVSS) in existing literature~\cite{secret_sharing_survey} only provides verifiability of the generation and dissemination of secret shares. However, in timed-release cryptography, verifiability of the revealed secret shares is also required to ensure the reconstructed secret is correct. This novel secret sharing scheme introduces such a property named reveal-verifiability, that provides public verifiability in the secret reconstruction stage.  

The main contributions of this paper are listed below:
\begin{itemize}
    \item A novel secret sharing scheme with reveal-verifiability to provide verifiability in secret reconstruction and efficient communication under a public communication model. 
    \item A secure timed-release cryptography system is constructed using smart contract and the proposed secret sharing scheme. The same architecture is applicable to achieve secure conditional information reveal systems in general.
    \item An open-source software prototype~\cite{source_code} of the proposed timed-release cryptography system that is deployed and tested across the globe. 
    \item A complete evaluation portfolio that includes analytical results, a validation of robustness in the Tamarin Prover and a real-world performance evaluation in a testnet processing more than 1,000 transactions.
    \item A real-world e-voting scenario to showcase the vulnerabilities of elections by strategic voting addressed by the proposed system to provide strong fairness to e-voting with highly secured ballots.
\end{itemize}

The rest of the paper is structured as follows: Section 2 reviews the related work in conditional information reveal systems and timed-release cryptography. Section 4 presents a smart-contract-based timed-release cryptography system. Section 3 presents a new secret sharing protocol that secures the proposed system. Section 4 presents a smart-contract-based timed-release cryptography system. Section 5 applies the proposed architecture to secure general conditional information reveal systems using smart contracts and secret sharing. Section 6 shows the advantages of the proposed system compared to the existing methods. Section 7 showcases the applicability of the proposed system in an e-voting scenario and Section 8 concludes the paper with an outline of future work.  

\section{Related Work}
The focus of this section is placed on existing timed-release cryptography systems. This section also provides the necessary background in related cryptographic primitives based on which the proposed timed-release cryptography systems are designed.

\subsection{Timed-Release Cryptography Systems}
Among various conditional information reveal systems, timed-release information methods have gained attention due to their potential applicability in real-world scenarios such as auctions~\cite{time-lock-auction} and voting~\cite{timelock_voting_application}. However, practical implementation of these methods has been challenging, primarily in balancing security and performance requirements. 

Existing research in the field is known as time lock encryption or timed-release cryptography~\cite{rivest_time-lock_1996}. These systems involve clients encrypting messages and secret holders managing the release of encrypted messages at a client-specified time. Cryptographic literature has identified two main approaches introduced below to building such systems: the time lock puzzles approach and the secret holder approach. 

\subsubsection{The Time Lock Puzzles Approach}
In the time lock puzzles approach, the answer to the computational puzzle serves as the decryption key to the message~\cite{liu2018build, hutchison_time-lock_2011}. The computational difficulty of the puzzle is set to adjust the expected puzzle-solving time given an estimation of available computational resources. 

One advantage of this approach is that it eliminates the need for trust in keeping the message secret before the client-specified decryption time. As a result, several studies have explored time lock encryption using time lock puzzles. However, related proposals~\cite{syverson1998weakly, malavolta2019homomorphic, tardis, tibouchi_astrolabous_2021, mao2001timed, Chvojka2021Versatile, loe2022tide} face a critical challenge in ensuring that the decrypter does not exploit additional computing power to expedite decryption, making them impractical. 

In recent years, the emergence of blockchain technology has advanced research in constructing time lock puzzles. The proof-of-work consensus mechanism~\cite{bitcoin} offers a new way to construct time lock puzzles with more stable puzzle-solving times. Inspired by proof-of-work blockchains, researchers have studied the challenge of dynamically controlling puzzle difficulty by finding a ``computational reference clock"~\cite{liu2018build}. This clock measures the mapping between computational difficulty and real-world time. For example, Bitcoin is designed to map a puzzle to ten minutes of real-world time by adjusting its difficulty according to the available computational resources. Various methods are proposed in existing work to convert the proof-of-work consensus to a time lock puzzle system~\cite{lai2019fully,liu2018build, chae2020practical, jaques2021time, lai2023lattice, abram2024time}. 

Existing works try to produce a more accurate mapping between puzzle difficulty and the required time to solve the puzzle, for example, by using verifiable delay functions (VDF) to create ``proof of sequential work" systems~\cite{jaques2021time, lai2023lattice, abram2024time} that reduces the randomness involved in the puzzle solving time. However, to the best of our knowledge, there are no studies to improve the puzzle difficulty adjustment mechanism, which is found to be problematic in the proof-of-work consensus when requiring a stable puzzle-solving time. 

\begin{figure}[h!]
    \centering
    \includegraphics[width=9cm]{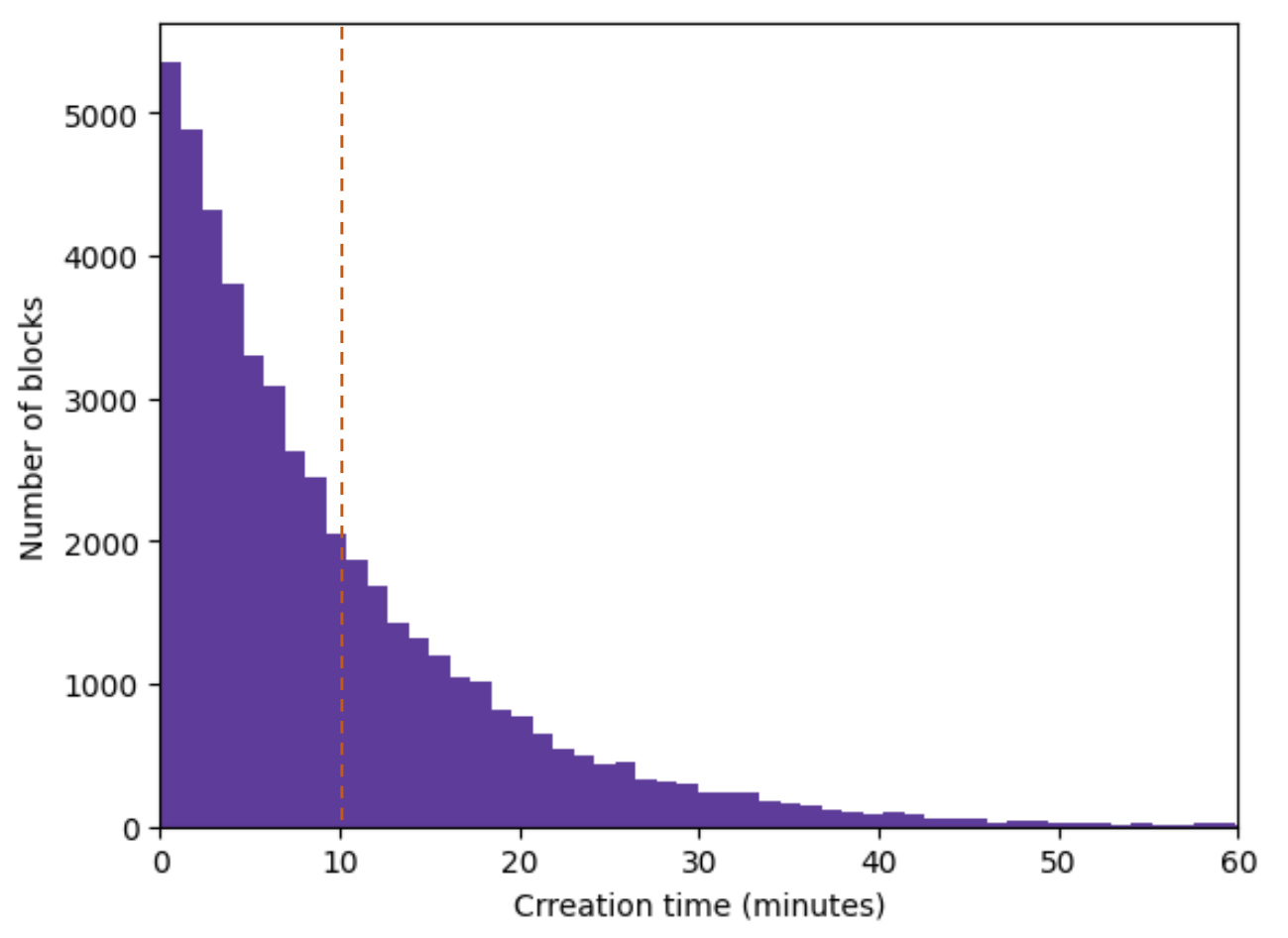}
    \caption{Bitcoin block time distribution from July 2022 to July 2023}
    \label{fig:block_creation_time}
\end{figure}
As the largest computational reference clock in the world that aims to solve a puzzle every ten minutes, the reality is far from its target. According to the Bitcoin block data from July 2022 to July 2023 provided by Google BigQuery~\cite{google_bitcoin_dataset}, although the average block time is computed as around 9.8 minutes, which is close to the target of 10 minutes per block, Figure \ref{fig:block_creation_time} indicates that creation time for individual blocks is not stable. The chance of a new block being created in 1 minute is more than two times larger than being created in exactly 10 minutes. Although with recent works on proof of sequential work, a better distribution can be achieved, it is still difficult to have a robust and dynamic difficulty adjustment mechanism to adjust the puzzle difficulties with the available computing powers. Current proof-of-work consensus adjusts puzzle difficulties based on the average block time of the previous 2016 blocks. When applied to time lock encryption, a malicious party can decide to use less or more computing power to slow down or speed up solving a particular puzzle to affect specific time lock messages with little effect on the difficulties of subsequent puzzles.

Additionally, clients have limited flexibility in choosing decryption times due to the alignment of puzzle-solving intervals with block creation times. Furthermore, time-lock puzzles inherently require significant computational resources, making them energy-inefficient. As energy consumption becomes a growing concern, blockchain consensus mechanisms are transitioning to more efficient and environmentally friendly alternatives. Therefore, there is a need to develop time lock encryption methods that are more energy-efficient.

\subsubsection{The Secret Holders Approach}
The other way is to make use of independent third parties to keep the secret. These parties hold decryption keys for clients and release them at the client-specified decryption time. This method relies on secret holders and provides a more accurate decryption time when the holders are honest and active. Therefore, efforts in existing work are put into securing the network from adversarial parties.

Centralized designs~\cite{timothy_timed-release_1993, di1999conditional, boneh2005hierarchical, boneh2000timed, chalkias2007improved, cheon2008provably, chan2005scalable} rely on one independent party acting as a secret holder. These methods focus on cryptographic-level security under the assumption of the honesty and activeness of the single secret holder. Security and robustness at a system level are not considered, such as potential single point of failure, and high level of trust required for the secret holder. The decentralized methods~\cite{bellare1997verifiable, rivest_time-lock_1996, rabin2006time, drand, cathalo2005efficient, celebi2018kimono, ning2018keeping, li2018decentralized, li2021silentdelivery, bacis2021told, ID-based_TRE} utilize threshold cryptography to provide fault tolerance while involving multiple secret holders to guard a secret. 

Although trust is distributed to multiple holders in decentralized solutions, there are yet security problems in early secret holder proposals. The first problem is the lack of reliability due to a lack of incentives. The question of ``why should clients trust the honesty of secret holders" is not addressed in the early proposals. Nothing stops secret holders from cheating. The second problem is data availability and authenticity. The need for ``public bulletin board"~\cite{rabin2006time} or ``publicly available location" to store data~\cite{rivest_time-lock_1996} is documented in existing proposals, but the robustness and authentication are not explicitly addressed. 

The development of blockchains and smart contracts addresses the challenges of incentives, data availability and authenticity. Recent proposals~\cite{celebi2018kimono, ning2018keeping, li2018decentralized, li2021silentdelivery, bacis2021told, ID-based_TRE} use smart contracts that integrate cryptocurrencies to provide incentives for secret holders, and ensure secret shares are immutable and available after revealing. However, none of these proposals provides reveal-verifiability. The definition of reveal-verifiability for secret sharing schemes is presented below. 

\begin{definition} \label{def: reveal-verifiability}
\textit{Reveal-verifiability}: Let \( k \) be a secret distributed among \( n \) participants using a secret-sharing scheme, and let \( R \) represent the reconstruction transcript, which includes the \( t \) pieces of information used for reconstructing the secret. The reconstruction process is defined by a deterministic reconstruction function \( \mathcal{R} \):
\[
k^* = \mathcal{R}(R),
\]
where \( R \) is the reconstruction transcript comprising the \( t \) secret shares used for reconstruction, and \( k^* \) is the output of the reconstruction function.

Then, a secret-sharing scheme satisfies reveal-verifiability if there exists a deterministic verification function \( \mathcal{V} \), accessible to all participants, such that:
\[
\mathcal{V}(R) =
\begin{cases}
\text{true}, & \text{if } \mathcal{R}(R) = k, \\
\text{false}, & \text{otherwise}.
\end{cases}
\]
\end{definition}

\paragraph{The reveal-verifiability problem}
To ensure a secret share revealed by a secret holder is correct, certain information as the witness is required. For example, Bacis et al.~\cite{bacis2021told} propose that a client publish the cryptographic hashes of the secret shares as witnesses. This allows the revealed shares to be hashed and compared against the witnesses for verification. However, this approach assumes the client is honest and does not account for the possibility of a malicious client publishing false witnesses. If a client publishes random hash values, the revealed secret shares are mistakenly deemed incorrect, leading to punishment of honest secret holders and compromising the integrity of the system. While malicious clients are modeled by Ning et al.~\cite{ning2018keeping}, but a solution to prevent such cases is not proposed. Other proposals~\cite{celebi2018kimono, li2018decentralized, li2021silentdelivery, bacis2021told, ID-based_TRE} similarly fail to address the threat of malicious clients, leaving them free to publish false witnesses without detection.

\paragraph{Comparison of Existing Secret Holder Approaches}
We provide a comparative analysis of the proposed method against existing secret holder methods in Table \ref{tab: security comparison}, focusing on several critical security features. Each approach is evaluated based on the following perspectives: resistance to single points of failure, incentives for honesty among participants, resilience to time source poisoning (changing secret holders local time by attacking the time servers), resilience to Sybil attacks, and reveal-verifiability under malicious clients. 

\begin{table*}[]
\centering
\caption{Security Comparison of Different Secret Holder Approaches}
\begin{tabular}{p{0.2\linewidth}p{0.12\linewidth}p{0.08\linewidth}p{0.15\linewidth}p{0.09\linewidth}p{0.16\linewidth}}
\hline
Method                                              & Single point of failure resistance & Incentives for honesty & Time source poisoning resistance   & Sybil attack resistance & Reveal-verifiability under malicious clients \\ \hline
Single secret holder: \cite{di1999conditional, boneh2005hierarchical, boneh2000timed, chalkias2007improved, cheon2008provably, chan2005scalable}                                & No                                 & No                     & No                                 & Yes                     & Yes           \\
Decentralized regular supply: \cite{cathalo2005efficient, drand} & Yes                                & No                     & No                                 & No                      & Yes           \\
Decentralized on-demand (off-chain): \cite{rivest_time-lock_1996}         & Yes                                & No                     & No                                 & No                      & No            \\ 
Decentralized on-demand (smart-contract-based): \cite{celebi2018kimono, ning2018keeping, li2018decentralized, li2021silentdelivery, bacis2021told, ID-based_TRE} & Yes & Yes & No & Yes & No \\ \hline
Our approach                                        & Yes                                & Yes                    & Resists early decryption & Yes                     & Yes           
\end{tabular}
\label{tab: security comparison}
\end{table*}

Although the single secret holder approaches are resistant to Sybil attacks due to the centralized, permissioned nature of the secret holder, the most critical problem they suffer from is a lack of resistance to single points of failure. 

Decentralized regular supply approaches use multiple secret holders to periodically supply key pairs for encrypting and decrypting secrets. They improve upon the single secret holder's model by distributing the responsibility across multiple entities, thus enhancing resistance to single points of failure. However, they introduce opportunities for Sybil attacks by allowing any parties to become secret holders with no restriction. Moreover, they still do not provide incentives for honesty, nor are they designed to resist time source poisoning. 

The decentralized on-demand approach only provides services upon receiving clients' requests. The existing proposal~\cite{rivest_time-lock_1996} is closest to our approach as our approach also falls into this category. However, it does not offer incentives for honesty, resistance to time source poisoning, or Sybil attack resistance. Incentives and Sybil attack resistance are resolved in the smart-contract-based proposals~\cite{celebi2018kimono, ning2018keeping, li2018decentralized, li2021silentdelivery, bacis2021told, ID-based_TRE}, but there lack of discussion on more secure time references of decryption time. More importantly, they also fall short in terms of reveal-verifiability in the presence of malicious clients.

Presented in the remaining sections, our approach distinguishes itself by leveraging security features from smart contracts and incorporating secret sharing with reveal-verifiability under malicious clients, providing security features with flawless incentives that are not achieved in other existing approaches.

\subsection{Secret Sharing Schemes for Secret Holders Approach}

Secret sharing schemes, in particular Shamir's Secret Sharing~\cite{shamir1979share}, are the core of multi-secret-holder timed-release cryptography systems. 

Shamir's Secret Sharing splits a secret $s$ into $n$ shares, such that the secret can be reconstructed using $t$ out of $n$ shares, where $t$ is the threshold and $n$ is the total number of shares. This is achieved by constructing a polynomial with a degree of $t-1$. The client generates $t-1$ random numbers $C_1, ..., C_{t-1}$, along with $s=C_0$ as the coefficients of the polynomial $P$:
\[P(x)=C_0+C_1\cdot x^1+...+C_{t-1}\cdot x^{t-1}\].
The evaluations of $P$ at different points are shared with the secret holders, such that each secret holder $i$ possesses $(i, P(i))$, a unique point on $P$, as its own secret share. With any $t$ out of $n$ points, the polynomial can be reconstructed via Lagrange interpolation, thus revealing $s=P(0)$.

The original Shamir's Secret Sharing scheme~\cite{shamir1979share} lacks a mechanism to verify the correctness of the secret shares distributed to the secret holders. While verifiable secret sharing protocols \cite{chor1985verifiable, feldman_practical_1987} introduce verification in the distribution phase to ensure that holders receive valid shares, they do not address the need for verifiability in the reveal phase, where the correctness of the recovered secret must be confirmed. Recent secret sharing protocols for smart-contract-based timed-release cryptography systems introduce various variants of secret sharing, for example, Li and Palanisamy~\cite{li2021silentdelivery} proposed an interactive protocol that enables secret holders to select points on the polynomial to use as shares without revealing them to the client; Jiang et al.~\cite{ID-based_TRE} incorporate identity-based encryption into the secret sharing process. However, these variants still fail to achieve reveal-verifiability, allowing malicious clients to frame honest secret holders in the reveal stage.

Moreover, secret sharing schemes often require peer-to-peer communication channels to distribute secret shares. When applied to timed-release cryptography, clients send secret shares to secret holders and also the time to reveal the shares. It is hard to ensure each secret holder receives the correct secret share and the same reveal time via a peer-to-peer channel. Applying existing secret sharing schemes in a public communication channel results in significant communication overhead, given that clients are required to publish all encrypted shares to deliver a unique secret share to each secret holder. This results in a large message size of $n \cdot |s|$, where $n$ is the number of holders and $|s|$ is the size of a secret share. Each holder receives all $n$ encrypted secret shares, even though only one share is relevant to each holder.

Based on the unique requirements of reveal-verifiability and public communication channels to apply secret sharing in timed-release cryptography, a new cryptographic protocol is introduced in this work to fulfill these two requirements. 

\subsubsection{Required Background}
The cryptographic protocol proposed in this paper advances secret sharing schemes under a public communication channel to achieve verifiability of recovered secrets using the necessary cryptographic primitives here. 

\paragraph{Diffie-Hellman Key Exchange}
The Diffie-Hellman key exchange enables two parties to derive a shared secret. The security of the protocol is based on the hardness of the discrete logarithm, which means it is difficult to find $a$ given $g^a$ in a group where the discrete logarithm is hard. To share a secret between two parties, Alice and Bob, using the protocol, Alice generates a secret $a$ and sends $g^a$ to Bob, and Bob generates a secret $b$ and sends $g^b$ to Alice, where $a$ and $b$ are random secret elements generated by Alice and Bob, and $g$ is a public parameter. Alice and Bob both obtain $g^{ab}$ as their secret, while adversaries cannot obtain any information about the secret given $g, g^a, g^b$.

\paragraph{Bilinear Pairings}
Bilinear pairings create a field where the computational Diffie-Hellman problem is hard but the decisional Diffie-Hellman problem is easy~\cite{CDH}. That is, given $g^a$ and $g^b$, it is hard to find out what is $g^{a\cdot b}$ (computational Diffie-Hellman is hard), but it is easy to find out whether $g^c = g^a \cdot g^b$ holds (decisional Diffie-Hellman is easy). 

In a pairing-friendly elliptic curve~\cite{barreto2003constructing}, such as BLS12-381, one can construct a pairing $e$ that inputs an element on a field $G_1$ with a generator $g_1$ and an element on a field $G_2$ with a generator $g_2$, and outputs an element on $G_T = G_1 \times G_2$, which is a multiplicative group of a field extension. The pairing $e$ is a bilinear map that comes with an important property: $e(a^c, b) = e(a, b^c)$, which is used to efficiently solve decisional Diffie-Hellman in these fields.

\section{Secret Sharing with Reveal-Verifiability on Public Communication Channel}
The proposed protocol in this section is not only applicable to the timed-release cryptography system. This protocol itself makes an independent contribution to cryptographic secret sharing techniques by achieving reveal-verifiability and can be applied to other scenarios where verifiable message recovery is required.

A vital requirement in designing the proposed system is to provide verifiability to the correctness of message decryption. This section presents a new cryptographic protocol to fulfill this requirement. The proposed protocol achieves linear communication cost when the number of participants increases, without the need for a peer-to-peer communication channel. The relevant math symbols are listed in Table \ref{tab: math symbol}. 

\begin{center} 
\begin{table}[]
\caption{Mathematical notations.}
\begin{tabular}{lp{7cm}}
\hline
Symbol & Meaning                                                           \\ \hline
$sk$   & Secret key                                                        \\ 
$pk$   & Public key                                                        \\ 
$t$    & Threshold number of secret holders required for decryption
\\ 
$n$    & Total number of secret holders                                    \\ 
$s$    & Secret share                                                      \\ 
$k$    & Symmetric key to encrypt and decrypt the message
\\ 
$m$    & Message plaintext                                                 \\ 
$c$    & Message ciphertext                                                \\ 
$g$    & Generator of a group                                              \\ 
$g_1$  & Generator of the first curve in BLS12-381                         \\ 
$g_2$  & Generator of the second curve in BLS12-381                        \\ 
$P$    & Lagrange polynomial                                               \\ 
$\alpha$ & Ciphertext of a polynomial evaluation                           \\ \hline
\end{tabular}
\label{tab: math symbol}
\end{table}
\end{center}

\subsection{Problem Modeling and Assumptions}
The objective of the cryptographic protocol is to distribute preserve confidentiality of a fixed-length information $k$. When applied to the timed-release cryptography system, $k$ is a symmetric key to encrypt and decrypt a time-sensitive message $m$. Two roles are involved in this protocol, including the client that owns $k$ and the $n$ secret holders that receive secret shares of $k$ denoted as $s_1, ..., s_n$. Here we define this required confidentiality property:

\begin{definition} \label{def:confidentiality}
    \textit{Confidentiality}: For any probabilistic polynomial-time (PPT) adversary \( \mathcal{A} \), the adversary's advantage in obtaining \( k \) fewer than \( t \) shares is negligible. That is:
    \[
    \Pr\big[\mathcal{A}(s_1, s_2, \ldots, s_{t-1}) \to k\big] \leq \epsilon,
    \]
    where \( \epsilon \) is a negligible function, \( s_1, s_2, \ldots, s_{t-1} \) are the shares available to the adversary, and \( t \) is the reconstruction threshold of the secret-sharing scheme.
\end{definition}

Unlike existing secret sharing protocols, since the proposed architecture is blockchain-based, it is required that clients and secret holders only communicate via sending permissionless blockchain transactions, i.e. in a publicly available broadcasting channel. Moreover, reveal-verifiability (Definition \ref{def: reveal-verifiability}) is an additional requirement compared to existing secret sharing methods. 

We make the following cryptographic and system level assumptions as the security foundations of the proposed protocol:
\begin{itemize}
    \item We assume the discrete logarithm problem is hard, that is, Let \( G_1 \) and \( G_2 \) be cyclic groups of prime order \( p \) with generators \( g_1 \) and \( g_2 \), respectively. Given \( g_1 \) and \( g_1^x \), it is computationally infeasible for a probabilistic polynomial-time adversary to compute \( x \in \mathbb{Z}_p \).
    \item At the system level, we assume at least a threshold $t$ out of $n$ secret holders are honest. 
    \item The randomly generated values, including private keys $sk_1, ..., sk_n$ from secret holders and the clients' random values $r$ (introduced below), are assumed to be secure and confidential.
    \item The broadcasting channel is assumed to be reliable such that all messages are eventually delivered to all parties. 
\end{itemize}

\subsection{Design}
The design of this protocol is broken down into three steps: (i) Finding the most efficient approach for the client to share with each holder a piece of secret information $s$ via a broadcasting channel. (ii) Connecting the secret shares to $k$, so that any $t$ different pieces of $s$ can be used to recovery $k$. (iii) Publicly verifying the correctness of the recovered $k$.

Diffie-Hellman key exchange allows a client to efficiently share with each holder a piece of secret share. As a setup, each secret holder is required to have an asymmetric key pair $(sk, pk)$, where the public key $pk=g^{sk}$ is known by clients and other holders. This is ensured in the secret holder registration process. Each holder must provide $pk$ and a digital signature of a message using this key pair to prove the procession of the corresponding $sk$. Moreover, each holder has a publicly known index $i$, given by the smart contract. The key pair of the $i^{th}$ holder is represented by $(sk_i, pk_i)$. By publishing the public keys of the holders, a client can send secret shares to all holders using a single short message, regardless of the number of holders. The public key of a holder can be utilized as the Diffie-Hellman shared message from the holder to the client. Specifically, the client can generate a random value $r$ and broadcast the value $g^r$. Consequently, each holder $i$ obtains a secret value $s_i = g^{r \cdot sk_i}$, where $sk_i$ represents the private key of holder $i$. The confidentiality of $s_i$ is ensured by the computational difficulty of the Diffie-Hellman problem. 

With secret shares communicated through Diffie-hellman, bilinear pairing can be adopted to achieve verifiability of the secret shares, i.e. $s_i$ revealed by holder $i$ is correct. Note that Diffie-Hellman key exchange is secure against attackers that aim to find out the exchanged keys in any fields where discrete logarithm is hard to compute, therefore, it is applicable in BLS12-381, a pairing-friendly curve, on which discrete logarithm is hard. Therefore $s_i$ should be computed by $s_i = g_1^{r\cdot sk_i}$ and the client should also broadcast $g_1^r$. On BLS12-381, $s_i$ can be verified by comparing $e(pk_i, g_2^r)$ and $e(s_i, g_2)$. This is correct because $e(pk_i, g_2^r) = e(g_1^{sk_i}, g_2^r) = e(g_1^{r\cdot sk_i}, g_2)$. To perform this check, a client only needs to publish $g_2^r$, regardless of the number of holders.

The next step is to establish a connection between these secret shares and $k$, such that any $t$ out of $n$ secret values can recover $k$. Similar to the existing secret sharing schemes, we hide $k$ in the evaluation of a $t-1$ degree polynomial. This polynomial $P$ is constructed by interpolating the points $(1, s_1), ..., (t-1, s_{t-1})$ along with $(0, k)$. The remaining shares, $s_t, ..., s_n$ form a relation with $k$ by transforming them into $P(t), ..., P(n)$. The transformation is achieved with encryption and decryption. Regard $P(t), ..., P(n)$ as messages to encrypt, $s_t, ..., s_n$ are the symmetric keys to encrypt these messages, the ciphertexts denoted as $\alpha_t, ..., \alpha_n$ can be obtained and published. This means the owner of $s_i$, holder $i$, can use $s_i$ to decrypt $\alpha_i$ to obtain the evaluation of $P$ at $i$. Since $s_i$ is a one-time key, the encryption scheme is essentially a one-time pad. As a concrete example, the exclusive OR operation can be used as the encryption and decryption method, i.e. $\alpha_i = P(i) \oplus s_i$. Therefore, computing and publishing $\alpha_t, ..., \alpha_n$ allows holder $t$ to holder $n$ to obtain $P(t), ..., P(n)$. To conclude, the secret share $s_i$ gives each secret holder a distinct evaluation on $P$. For holder $1$ to holder $t-1$, $s_i$ equals to $P(i)$; for holder $t$ to holder $n$, $s_i$ is the key to decrypt the ciphertext $\alpha_i$ to obtain the plaintext $P(i)$. 

\begin{figure}[h!]
\centering
\includegraphics[width=9cm]{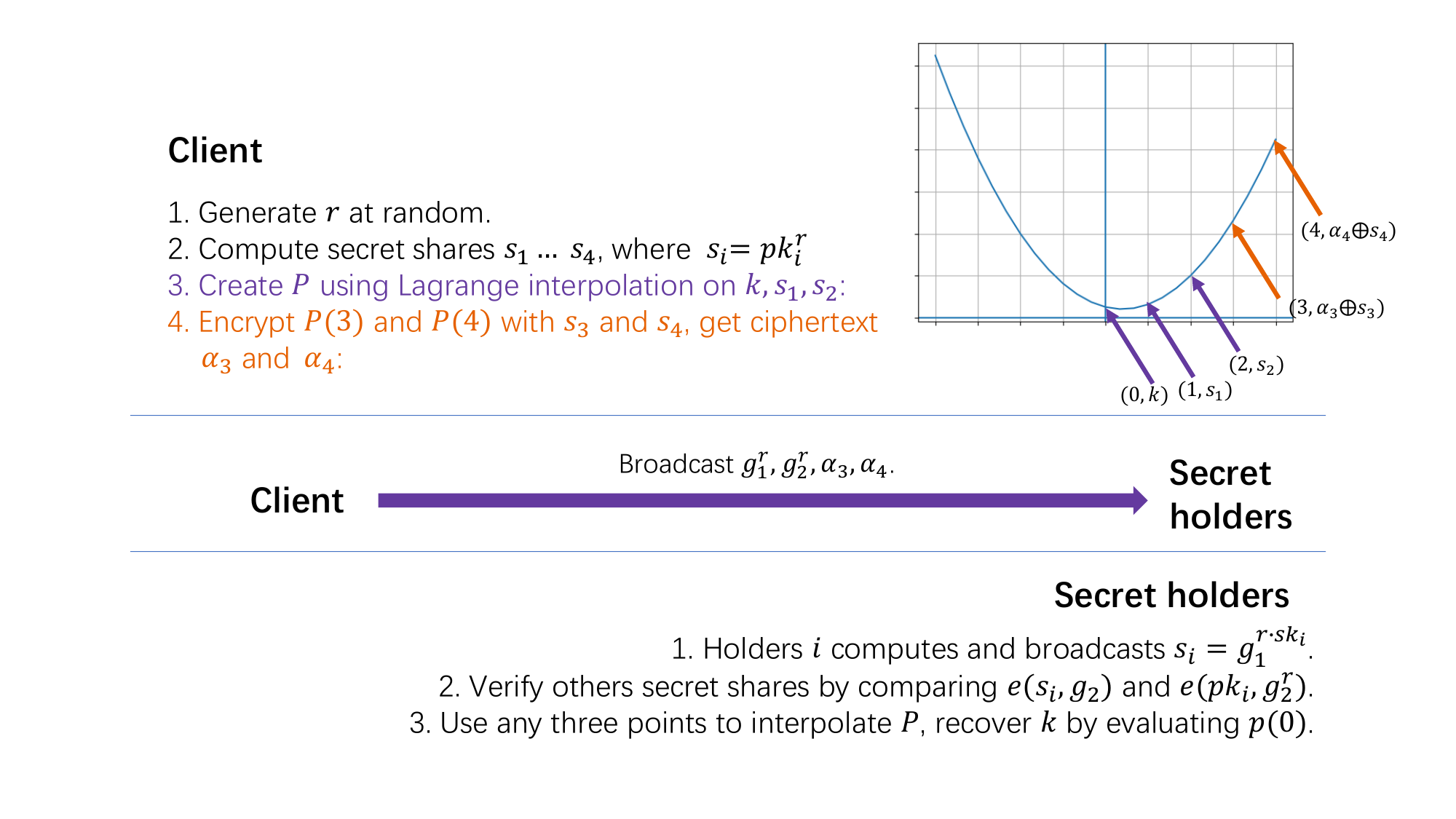}
\caption{An example of the proposed secret sharing method with four secret holders}
\label{fig:crypto_protocol_example}
\end{figure}

Figure \ref{fig:crypto_protocol_example} presents an example of a client sharing $k$ to four holders with a threshold of three holders to recover $k$. A second degree polynomial $P$ is generated by the client using $(0, k), (1, s_1), (2, s_2)$; $s_3$ and $s_4$ are used to encrypt and decrypt $P(3)$ and $P(4)$. 

The reveal-verifiability of the recovered secret is provided by the verifiability of each individual secret share from bilinear pairing evaluation. Given that the secret shares are verifiably correct and the $\alpha$s are public, anyone can use the same inputs on Lagrange interpolation to reproduce $k$. 

Note that this protocol is also a traditionally defined verifiable secret sharing in the sense that clients cannot communicate incorrect secret shares to secret holders, as the secret shares are all derived deterministically from the same value $g_1^r$. 

\subsection{Applying to Timed Release Cryptography}
Combining all components in the design, we go over the entire process of the proposed cryptographic protocol when it is applied to the timed-release cryptography system. Suppose the client has a message $m$ to encrypt for a period of time; through the published smart contract that stores information of the system, the client knows the public keys of the secret holders $pk_1, ..., pk_n$.

\noindent \textbf{Actions of Clients}
\begin{enumerate}
    \item The client generates two random values $(k, r)$. 
    \item The client uses $k$ to encrypt the message $m$ with symmetric encryption, denote the ciphertext as $c$. 
    \item The client computes secret shares $(s_1, ..., s_n) = (pk_1^r, ..., pk_n^r)$. 
    \item The client computes the polynomial $P$ using Lagrange interpolation on points $(0,k), (1, s_1), (2, s_2), ..., (t-1, s_{t-1})$. 
    \item The client computes $\alpha_t = P(t) \oplus s_t, ..., \alpha_n = P(n) \oplus s_n$, which are the ciphertexts of $P(i)$ for $t \leq i \leq n$.
    \item The client computes two more values: $g_1^r$ for the secret holders to derive secret shares, and $g_2^r$ to verify shares revealed by the secret holders.
    \item Broadcast a request to all holders including the ciphertext of the message $c$; decryption condition $time$; $g_1^r$; $g_2^r$, and $\alpha_t, ..., \alpha_n$.
\end{enumerate}

\noindent \textbf{Actions of Secret Holders}
\begin{enumerate}
    \item When receiving a request from a client, get the secret share $s_i = pk_i^r = g_1^{r\cdot sk_i}$.
    \item Wait until the client specified $time$ is reached and publish $s_i$.
    \item Verify secret shares submitted by other secret holders by evaluating whether $e(s_i, g_2) = e(pk_i, g_2^r)$ holds.
    \item Once $t$ pieces of $s_i$ are received and verified as correct shares, the secret holders recover $k$ by evaluating $P(0)$ using $t$ points on $P$. The $t$ points are obtained as follows: for $s_1, ..., s_{t-1}$, the point is $(i, s_i)$; for $s_t, ..., s_n$, the point is $(i, s_i \oplus \alpha_i)$.
    \item Decrypt the ciphertext $c$ with $k$ and publish $m$.
\end{enumerate}

\section{A Smart-Contract-based Timed-Release Cryptography System}
This section illustrates a timed-release cryptography system using the decentralized conditional information reveal architecture. Sharing a similar basic idea with other smart-contract-based constructions~\cite{celebi2018kimono, ning2018keeping, li2018decentralized, li2021silentdelivery, bacis2021told, ID-based_TRE}, Our system additionally achieves higher reliability and security by making uses of the blockchain clock and achieving reveal-verifiability to prevent malicious clients from framing honest secret holders.

\subsection{The Smart-Contract-based Architecture}
\begin{figure}[h!]
\centering
\includegraphics[width=9cm]{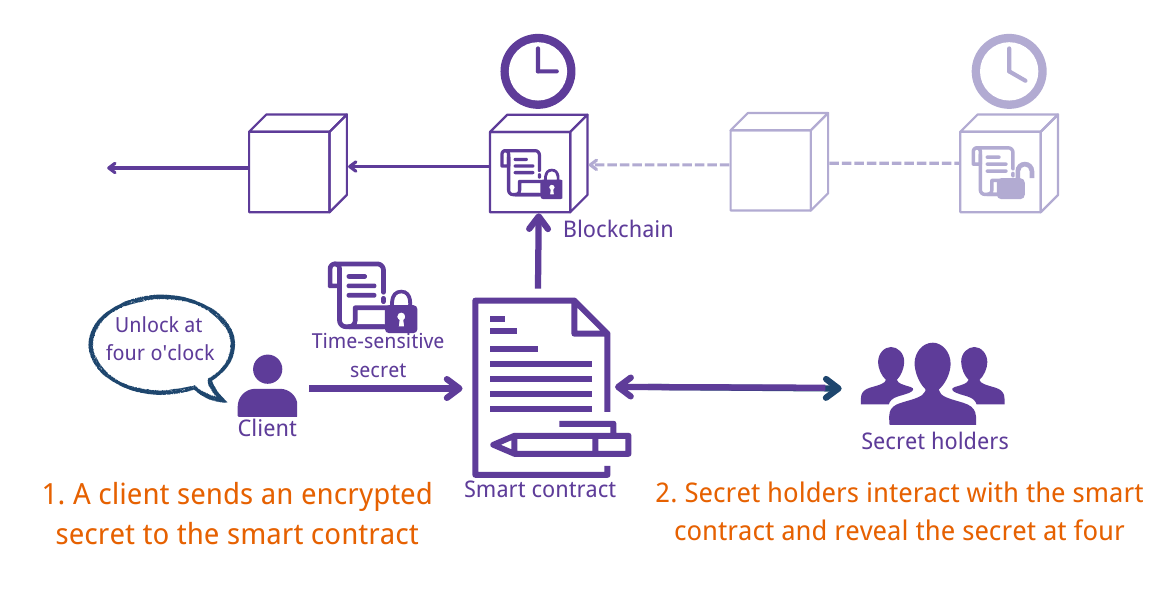}
\caption{The high-level process of sending a timed-release message}
\label{fig:overiew_figure}
\end{figure}

Figure \ref{fig:overiew_figure} shows a high-level overview of the system with an example. To use the system, clients submit encrypted messages to the smart contract, and secret holders store the messages in encrypted form. The holders then decrypt and publish the messages on the blockchain at the specified decryption time determined by the client. Clients are not responsible for decrypting secret messages, allowing them to exit the process after submitting the encrypted message. For example, a client can send an encrypted message at 3 o'clock and ask it to be committed at 4 o'clock. 

\subsection{Improved Clock Reliability with On-Chain Timestamps} \label{sec: 15s}
The system employs two references of time to verify time conditions. The first source is the synchronized time with a centralized global clock through the network time protocol~\cite{networktimeprotocol}. This is a built-in functionality of modern computers~\cite{linux_ntpd_doc}, providing high-precision time references. Secret holders within our system utilize this centralized clock to verify time conditions specified by clients. Notably, numerous reliable time servers are readily available on the Internet, further enhancing the robustness of this time reference.

In addition to the centralized global clock, the system uses the power of the decentralized blockchain clock as an auxiliary time source. While distributed systems such as blockchains do not achieve perfect clock synchronization, they offer a resilient and available time reference with bounded drift given the safety and liveness of the blockchain~\cite{blockchain-time-accurate, blockchain_time_study, pow_clock_analysis}. This blockchain clock is grounded in the concept of block time, which is manifested as timestamps embedded within block headers. 

The blockchain clock plays a crucial role as a fail-safe mechanism in our system to ensure that secrets are not revealed earlier in time. Tzinas et al.~\cite{blockchain-time-accurate} have shown that timestamps produced from blockchains have high accuracy. Take Ethereum as an example, the network mandates validators to synchronize their time with the network upon joining~\cite{ethereum_yellowpaper, geth_doc}. In cases where a validator's local time deviates significantly from the network time, it results in isolated blocks that are not accepted by the majority of the network. Implementation-wise, it only accommodates a 15-second forward time difference between a validator's local clock and the clock of the block proposer~\cite{geth_codebase}, which means a malicious block proposer can only advance the Ethereum network's time by a bounded 15-second window. This provides additional resistance against early message decryption. In the extreme case where a majority of secret holders' local clock failed and an attacker is chosen as the block proposer involving message decryption, Ethereum still serves as an effective tool in preventing the 15-second onward advanced disclosure of the message. By incorporating these two complementary time references, the proposed system strikes a balance between precision and robustness, ensuring the secure and timely release of confidential information.

\subsection{Verifiability of Revealed Information} 
To ensure the messages are reconstructed correctly by the secret holders and prevent malicious behavior from both holders and clients, the proposed secret sharing protocol is adopted. 

While smart contracts possess the capability to verify secret shares on-chain, this process is offloaded to holders for off-chain execution to reduce the gas costs incurred in the on-chain computation. A parallel in design can be drawn to the optimistic rollup, a well-known blockchain scaling methodology~\cite{optimistic_rollup}. Mirroring the fault proof mechanisms in an optimistic rollup, in the proposed system, a secret share submission enters a provisional state for a set duration, for example, an hour. This interim period allows holders to verify the submission and possibly raise disputes. If unchallenged within this window, the secret share gains validation and the submitter is rewarded.

\subsection{Example Workflow}
\begin{figure}
    \centering
    \begin{tabular}{@{}c@{}}
        \includegraphics[width=1\linewidth]{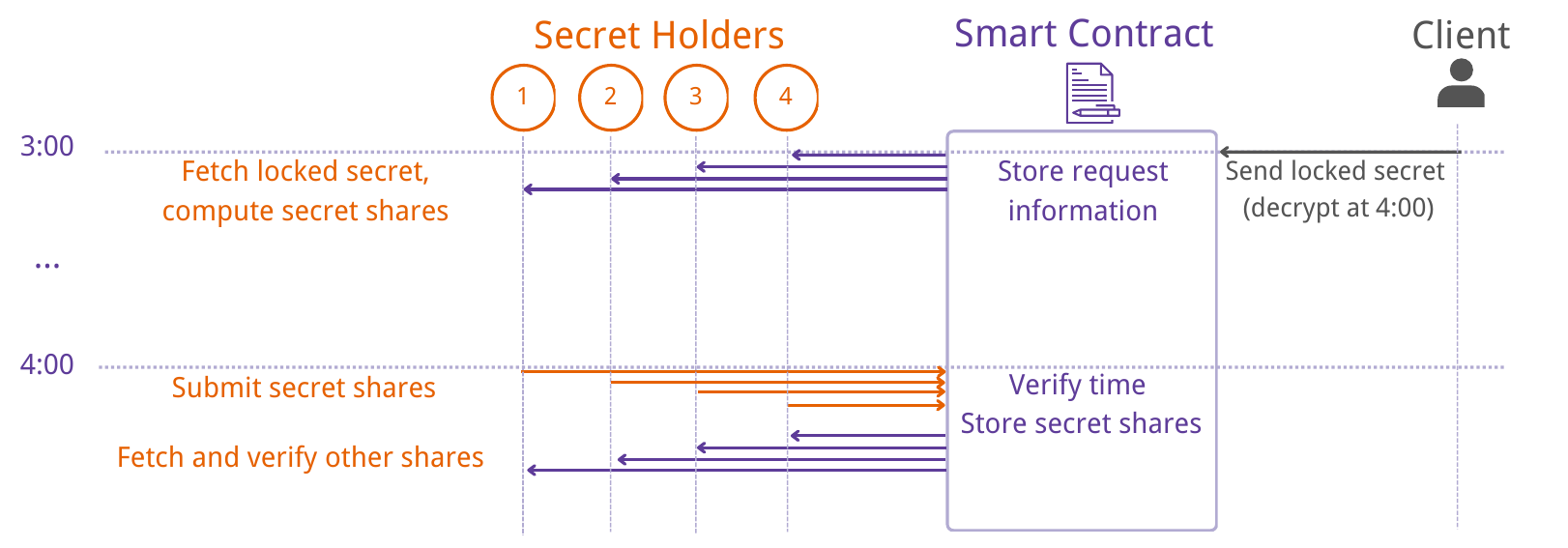} \\[\abovecaptionskip]
        \small (a) 4 honest secret holders
        \\
    \end{tabular}
    \begin{tabular}{@{}c@{}}
        \includegraphics[width=1\linewidth]{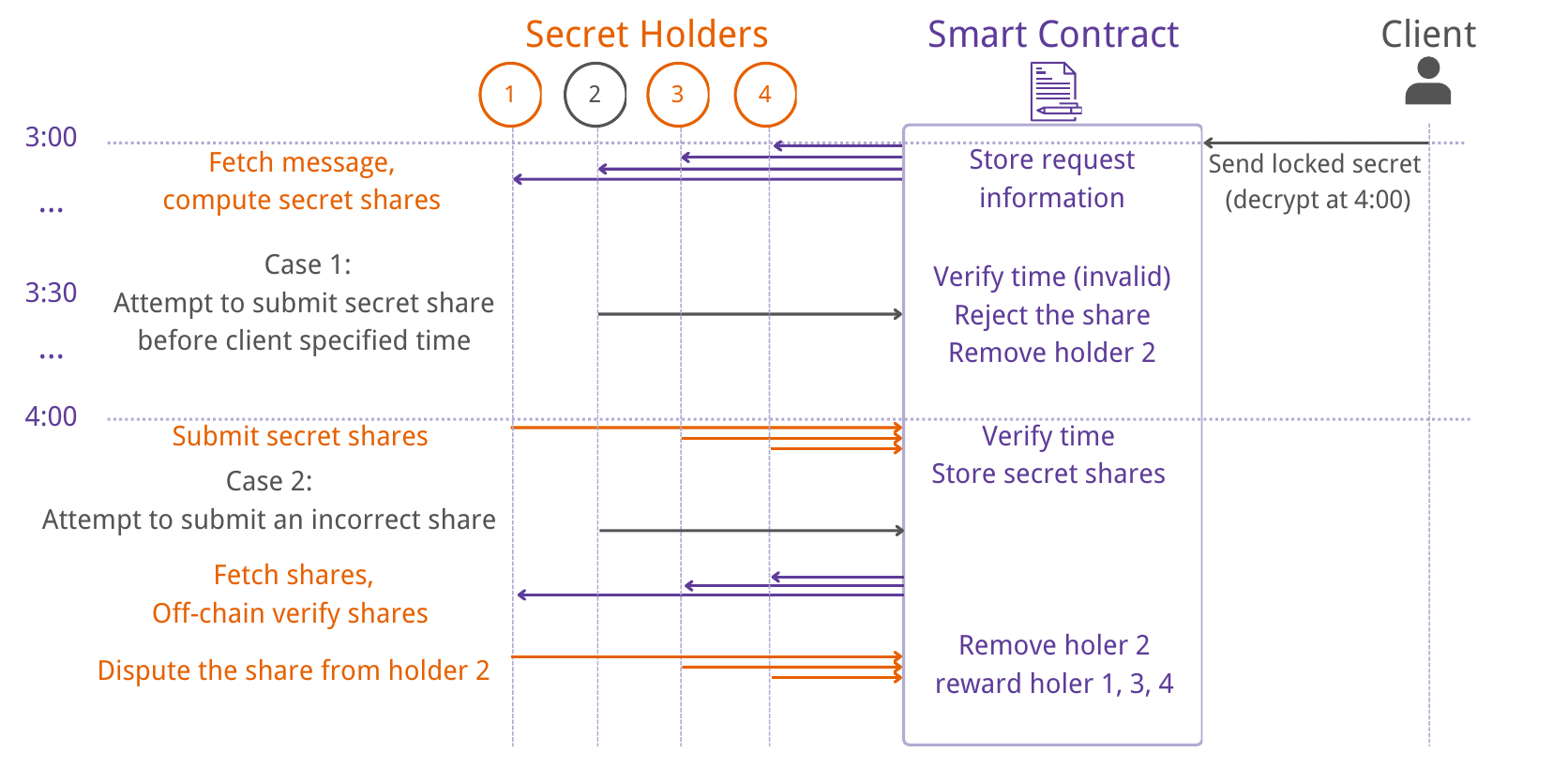} \\[\abovecaptionskip]
        \small (b) 3 honest holders and 1 adversarial holder
        \\
    \end{tabular} 
    \caption{Example workflow in the view of on-chain activities}\label{fig:workflow}
\end{figure}

Combining all the design components, here we introduce as an example the workflow of the network on Ethereum. The example system consists of four independent parties as holders. The four parties deposit Ethers to the smart contract and register as holders. At the same time, they setup Ethereum nodes or use APIs from Ethereum node providers to monitor events of the smart contract. Figure \ref{fig:workflow}a and \ref{fig:workflow}b illustrate two scenarios, when all holders are honest and when a holder is adversarial.
\\ \\

\noindent

In the proposed system, the process begins when a client sends a timed-release message request to the smart contract. Upon receiving the request, the holders individually compute their own secret shares and wait until the client-specified decryption time to submit their shares to the smart contract.

To ensure the integrity of the system, if a member attempts to submit their shares before the decryption time specified by the client, it is automatically identified as a dishonest holder by the smart contract. In such cases, these holders are removed from the list of eligible holders and lose their deposit.

Upon reaching the client-specified decryption time, honest holders proceed to submit their shares to the smart contract, signifying the reveal of the secret. Subsequently, holders retrieve and locally verify the shares submitted by peer holders. If no disputes are raised, the smart contract allocates rewards to the first three holders who successfully submit their shares. Despite submitting a correct share, the fourth holder does not receive a reward, as its contribution is redundant for the secret reveal. This reward strategy incentivizes timely submissions of secret shares.

\section{Towards Secure Conditional Information Reveal Systems}
Taking a step back from timed-release cryptography, this section introduces a paradigm to show that smart contract platforms and the proposed secret sharing scheme also apply to generic conditional information reveal systems with key security advantages. 
 
\subsection{Security Challenges}
While existing conditional information reveal systems exhibit certain similarities, their design is confronted by distinct challenges. For instance, as a widely used identity-based conditional information reveal system, a foundational presumption in prevailing single-sign-on (SSO) systems is the implicit trust required between service providers and clients towards authentication service providers. Consequently, its centralized architecture introduces vulnerabilities, most notably, a single point of failure. In the realm of location verification, it is challenging to identify a method that is simultaneously reliable, efficient, and precise for positioning. Despite the prevalence of conditional information reveal systems, a systematic approach to identifying the challenges they face is absent. 

In summarizing the core components of conditional information reveal systems, we identify that the key security requirements for conditional information reveal systems include:
\begin{enumerate}
    \item Reliable communication channel: Establishing a reliable communication channel is crucial to ensure the authenticity of messages exchanged between clients and information holders.
    \item Reliable condition check: Some conditions may require additional facilities, devices, or protocols. For example, a location-based conditional information reveal system may need specific positioning devices. In a decentralized environment involving multiple holders, reaching a consensus on whether a condition is met becomes more challenging.
    \item Verifiable reveal: The correctness of the revealed information shared by holders should be verifiable by other parties to ensure that clients' requests are accurately served. 
    \item Activeness and honesty of information holders: The above three challenges are related to the execution process of the systems, however, to establish such systems, it often requires benefits to encourage participation of honest holders. 
\end{enumerate}

\subsection{Decentralizing Conditional Information Reveal Systems}
Considering the smart-contract-based proposals in time-based conditional information reveal system~\cite{celebi2018kimono, ning2018keeping, li2018decentralized, li2021silentdelivery, bacis2021told, ID-based_TRE} in a generalized view, this architecture can in fact be applied to all types of conditional information reveal systems. The only difference is that the secret holders require different methods to check if the required conditions are met.

Involving multiple secret holders through secret sharing and communicating through smart contracts improves system reliability. It provides a reliable communication channel where data authenticity and availability are ensured. Combining incentive mechanisms with cryptocurrencies and the proposed secret sharing protocol that provides reveal-verifiability, honest and active secret holders are always incentivized to secure the protocol.

\begin{figure}[h!]
\centering
\includegraphics[width=9cm]{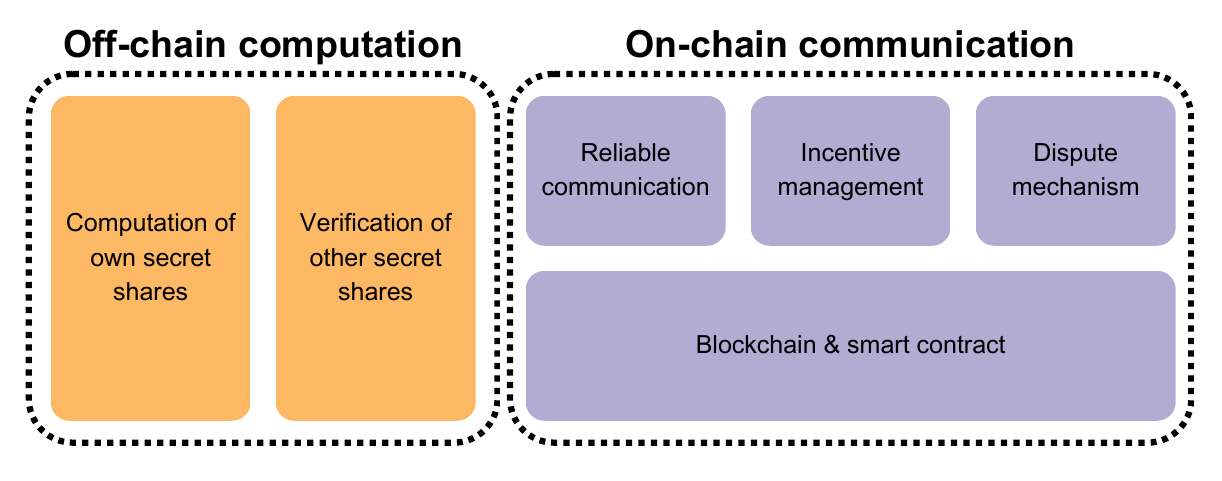}
\caption{Modules in the decentralized architecture for conditional information reveal systems using smart contract and secret sharing}
\label{fig:CIS_architecture}
\end{figure}

Figure \ref{fig:CIS_architecture} illustrates the modules of the smart-contract-based conditional information reveal architecture. Computation and verification of secret shares can be executed locally by holders to reduce on-chain computation costs; holders communicate and self-organize the network through a smart contract. This generic architecture provides adaptability to different conditional information reveal systems, accommodating various information sharing conditions. 

\section{Security Analysis}
Going back into the proposed timed-release cryptography system, this section outlines an attack model for the proposed system and provides formal proof of security within this framework. Moreover, a comparative analysis of different secret holder approaches is provided to demonstrate the advantage of the proposed system on critical security features.

\subsection{Attack Model}
Three types of adversaries are taken into account. 

\begin{itemize}
    \item \textit{External adversaries} are malicious parties that aim to breach confidentiality (see Definition \ref{def:confidentiality}) using public knowledge.
    \item \textit{Malicious secret holders} are parties that aim to breach confidentiality using public knowledge along with their secret shares. We assume at most $t-1$ secret holders are considered malicious. This assumption can be upheld through various measures, tailored to specific application contexts. For instance, in a voting scenario, voters themselves could act as secret holders, while in more general applications, monetary incentives via cryptocurrency could encourage honest behavior. Under these conditions, whether a minority of malicious secret holders could compromise the secrecy of a time lock message is assessed. 
    \item \textit{Malicious clients} are parties that aim to breach reveal-verifiability (see Definition \ref{def: reveal-verifiability}) of the protocol. The system fails if clients can falsely accuse honest secret holders of submitting incorrect shares. In this case, honest secret holders are punished and disqualified even if they follow the protocol. Therefore, the proposed protocol must ensure reveal-verifiability in the presence of malicious clients.
\end{itemize}

\subsection{Confidentiality}
Three lemmas are provided to prove the confidentiality of the proposed protocol. Lemma \ref{lemma: tamarin} and \ref{lemma: any_external} combined is a proof by induction showing that confidentiality holds against \textit{external adversaries} regardless of the number of secret holders in the system. Lemma \ref{lemma: adversarial_secret_holders} shows that confidentiality also holds against a minority of \textit{malicious secret holders}.

The Tamarin Prover~\cite{tamarin} is used to model the system with three secret holders, verifying the impossibility for \textit{external adversaries} to access two secret shares before their release. The Tamarin Prover is a tool for formal verification of cryptographic protocol. Unlike other commonly used cryptographic verification tools such as ProVerif~\cite{proverif} and CryptoVerif~\cite{cryptoverif}, Tamarin offers flexible support for adversary modeling. The blockchain communication model imposes restrictions on adversaries, preventing them from modifying public messages due to the security guarantees provided by the underlying blockchain. The Tamarin Prover accurately models this communication channel and verifies the authenticity of all public messages within our protocol. 

\begin{lemma}
    When $n=3, t=2$, For any PPT \textit{external adversaries} \( \mathcal{A} \) that have access to the public knowledge $(g_1, g_2, pk_1, pk_2, pk_3, g_1^r, g_2^r, \alpha_{3})$ and at most one secret share $(s_i, i\in\{1,2,3\})$, the probability of knowing $k$ is negligible.
    \label{lemma: tamarin}
\end{lemma}

\begin{proof}
    Suppose \( \mathcal{A} \) can obtain $k$ with the public knowledge and an $s_i$ with non-negligible probability, \( \mathcal{A} \) must be able to get an additional secret share $s_j=pk_j^r$. Then \( \mathcal{A} \) must be able to get $r$ from $g_1^r$ or $g_2^r$ with non-negligible probability, that is, \( \mathcal{A} \) must be able to solve the discrete logarithm problem with non-negligible probability, which contradicts to our assumption. 
    
    The lemma is also successfully proven by the Tamarin Prover. Source code can be found in the open-source repository~\cite{source_code} of this project. 
\end{proof}

With this base case of three secret holders. We extend this proof to show that message secrecy holds for any size of secret holder. 

\begin{lemma}
\label{lemma: any_external}
    For any $n>3$, any PPT \textit{external adversaries} \( \mathcal{A} \) has negligible probability of obtaining any secret share $s_i$ before its respective holder submits it.
\end{lemma}

\begin{proof}
    For \( \mathcal{A} \) to access a secret share $s_i$ before holder $i$ submits it in a system with $n>3$ holders ($i < n$), it must obtain a secret share prematurely from $n-1$ holders. The additional knowledge provided by the $n^{th}$ holder is $pk_n = g_1^{sk_n}$ and $\alpha_n = g_1^{sk_n*r} \oplus P(n)$. Based on the hardness of discrete logarithm problem,  \( \mathcal{A} \) has negligible probabilities to obtain $sk_n$ and $r$, and therefore has negligible probability to obtain $s_i = g_1^{sk_i*r}$. Similarly, the adversary must also be able to prematurely access a secret share with $n-2$ holders. Therefore, the adversary must always be able to access a secret share without the presence of one holder. When the number of secret holders is decreased to $n=3$, the adversary cannot obtain a secret share as proven in Lemma \ref{lemma: tamarin}. Hence, an external adversary has a negligible probability of prematurely accessing a secret share before the holder submits it at any system size of $n$.
\end{proof}

We further provide proof that the secrecy of the message key $k$ holds in the present of \textit{malicious secret holders} below the threshold. 

\begin{lemma}
\label{lemma: adversarial_secret_holders}
    Any $t-1$ PPT \textit{malicious secret holders} has negligible probability of knowing $k$. 
\end{lemma}

\begin{proof}
    Suppose the collaborating \textit{malicious secret holders} can obtain $k$ before honest holders submit shares, they must be able to gain one extra secret share belonging to the honest holders. That is, they are able to get the secret share using the public knowledge $g_1, g_2, g_1^r, g_2^r, pk_1, ..., pk_n, \alpha_{t+1}, ..., \alpha_{n}$ along with their own secret keys. Since their own secret keys are randomly generated values, which are irrelevant to any other honest holders' secret share, they must be able to obtain the secret share using public knowledge. However, lemma \ref{lemma: any_external} shows that it is not possible to obtain a secret share using public knowledge, which contradicts the initial assumption and proves the lemma.
\end{proof}

\subsection{Reveal-Verifiability}
Shares published by honest secret holders should be verified as correct even if the encryption request is initiated by a \textit{malicious client}. Below we present possible actions of a \textit{malicious client} and show that reveal-verifiability is not violated.

Denote $a=g_1^r$ for secret holders to compute the share as $s = a^{sk}$ and another value $b=g_2^r$ for the public to verify secret shares. In this notation, the verification process is to check whether $e(pk, b) = e(s, g_2)$ holds. A \textit{malicious client} can attempt to violate the protocol by publishing $a$ or $b$ as any other values other than $a=g_1^r$ and $b=g_2^r$. 

\begin{lemma}
    There exists a deterministic verification function $\mathcal{V}$, accessible to all participants, that identifies $a=g_1^r$ and $b=g_2^{r'}$ for any $r\neq r'$.
\end{lemma}

\begin{proof}
    A deterministic verification function $\mathcal{V}$ is defined below:
    \[
        \mathcal{V}(a,b,g_1,g_2) =
        \begin{cases}
        \text{true}, & \text{if } e(g_1, b) = e(a, g_2), \\
        \text{false}, & \text{otherwise}.
        \end{cases}
    \]
    $\mathcal{V}$ always output true for any $r\neq r'$ when $a=g_1^r$ and $b=g_2^{r'}$.
\end{proof}

As a result, \textit{malicious clients} are not able to corrupt the verifiability of secret shares.

\section{Performance Evaluation}
The most fundamental requirement for a timed-release cryptography system is to decrypt messages at the time users specify. Therefore, decryption time deviation, computed as the difference between the expected decryption time and actual decryption time is measured in this section under various scenarios. Moreover, we provide an estimation of the energy consumption of the proposed system, showing its advantage in energy efficiency.

The following performance data of the proposed system are obtained from a geographically distributed testbed in which the secret holder servers are well distributed across East US, West US, UK, North Europe and Australia. Microsoft Azure and Google Cloud Platform are used as server providers. Each secret holder runs an Ubuntu 20.04 server with 2 vCPU of 2.1 GHz and 4GB RAM. This showcases the low hardware requirement of the proposed system, enabling a low-cost formation of a distributed environment in practice. The code run by the secret holders is accessible in the project open-source repository~\cite{source_code}. 

The smart contract is deployed on the Arbitrum Sepolia testnet~\cite{arbitrum_sepolia}, a layer-2 blockchain for testing decentralized applications with a fast block time of less than 1 second. The secret holder servers use public RPC node endpoints provided by Alchemy~\cite{alchemy} to communicate with the blockchain.

\begin{figure}[h!]
\centering
\includegraphics[width=9cm]{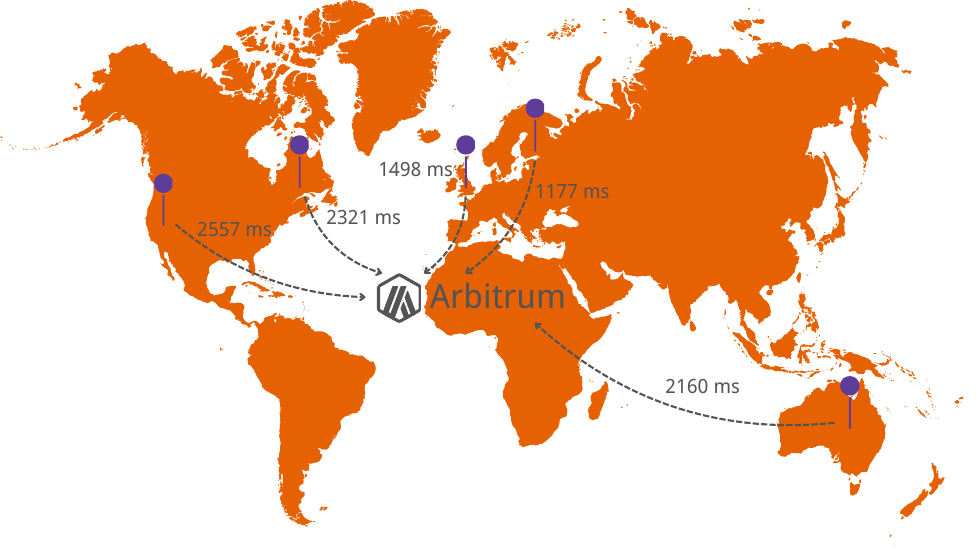}
\caption{Locations of the secret holder servers and their average latency of submitting secret shares to Arbitrum Sepolia}
\label{fig:geomap}
\end{figure}

Over 1,000 transactions were submitted to the blockchain during the experiment. As shown in Figure \ref{fig:geomap}, on average it takes around one to three seconds, depending on the physical location, for a secret holder to complete a transaction of submitting a secret share to the blockchain.  

\subsection{Decryption Time Stability}
Using the illustrated testbed, we compare the decryption time stability between time lock puzzle approaches and the proposed approach with decryption time deviation. For the time lock puzzle system, understood in the context of the Proof of Work consensus mechanism~\cite{lai2019fully,liu2018build, chae2020practical}, its decryption time deviation is represented using the block time data of the Bitcoin blockchain~\cite{google_bitcoin_dataset}. Given the large scale of the Bitcoin blockchain, it is considered the most stable time lock puzzle construction in reality. As the average time to produce one block in the Bitcoin blockchain is 10 minutes, it is also the minimum encryption duration of the system. An instance of the proposed system with ten secret holders is used to compare against the time lock puzzle approach. Requests with an encryption duration of ten minutes to one week are generated for both systems, with twenty dummy requests for each specified duration.

\begin{figure}[h!]
\centering
\includegraphics[width=9cm]{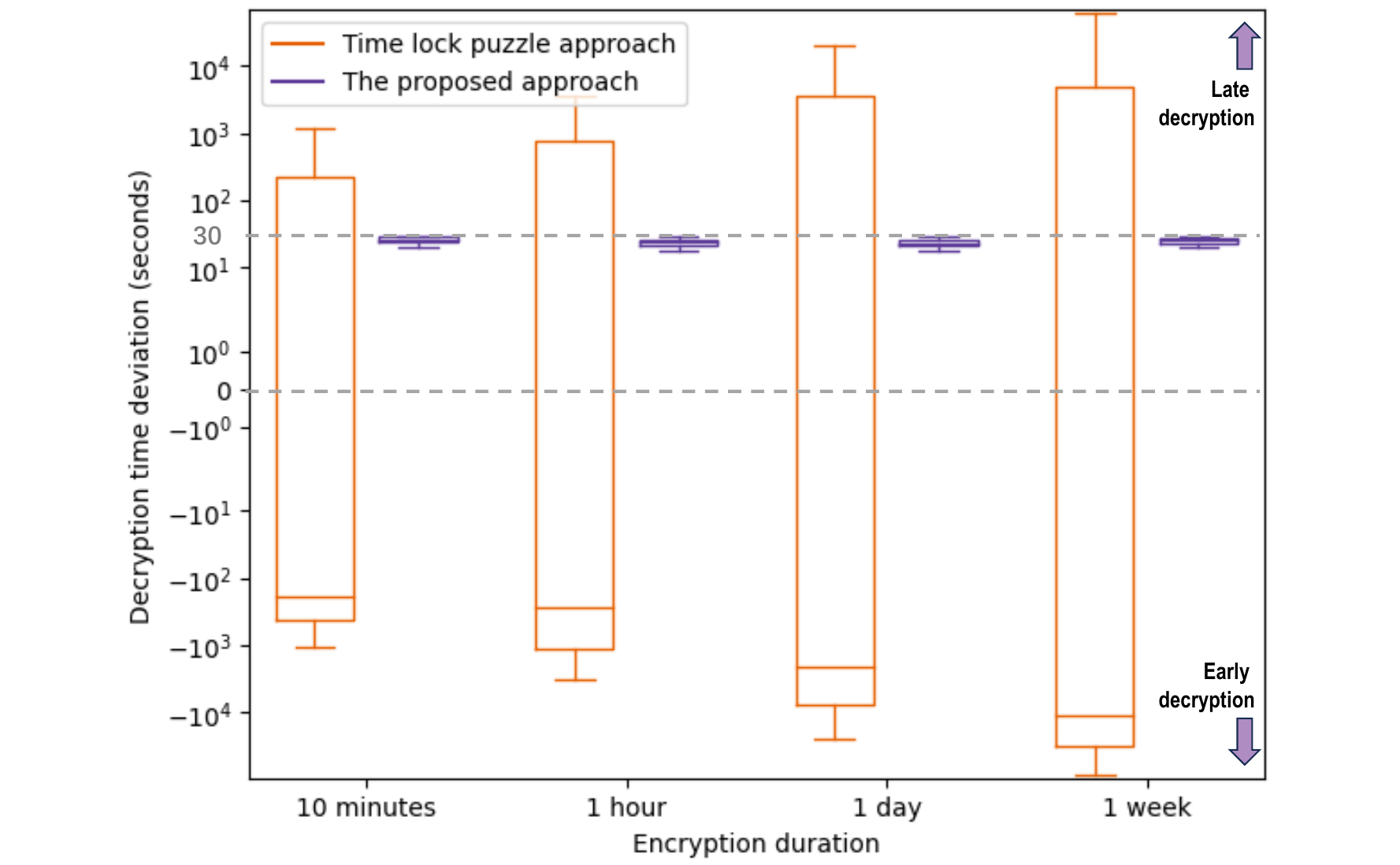}
\caption{Decryption time deviation of two systems in logarithmic scale}
\label{fig:performance_comparison}
\end{figure}

As indicated in Figure \ref{fig:performance_comparison}, the time lock puzzle approach exhibits unreliable decryption time. Its performance deteriorates with increasing encryption duration; for instance, a one-week encryption period results in an average decryption time deviation exceeding two hours. The result indicates two disadvantages of time lock puzzle systems. First, the long error bars illustrate the unstable puzzle-solving time. Second, the average lines at the early decryption area far from zero deviation show that the puzzle difficulty cannot be adjusted perfectly to achieve the desired average solving time. In the case of the Bitcoin blockchain, the hash rate of the network is in a strong increasing trend over time, while the increase of the puzzle difficulty does not strictly follow, resulting in an average block time shorter than expected, thus an accumulated deviation as the encryption duration increases. 

In stark contrast, our approach maintains a consistently low decryption time deviation from 18 seconds to 30 seconds, irrespective of the encryption duration. The nature of the secret holder-based approach provides a stable decryption time regardless of the encryption duration. A majority of the deviation is attributed to the local verification of secret shares on the secret holder servers (see Figure \ref{fig:scalability} for more details). Therefore, this deviation can be further reduced by employing more computationally powerful servers.

\subsection{Scalability}
An advantage of the time lock puzzle approach is that it is highly scalable regarding the computing power to solve puzzles as a result of the adaptability of puzzle difficulty; it is also highly scalable regarding the number of message requests, since all messages with the same decryption time are encrypted using the same puzzle. 

In the case of the proposed system, the latency in the decryption process comprises two elements: the time needed for holders to submit secret shares to the blockchain and the time to verify a sufficient number of secret shares. To understand the influence of scaling the number of secret holders on decryption latency, experiments of 3, 10, 20, 30, and 40 secret holders are conducted to process dummy time lock messages. 

\begin{figure}[h!]
\centering
\includegraphics[width=9cm]{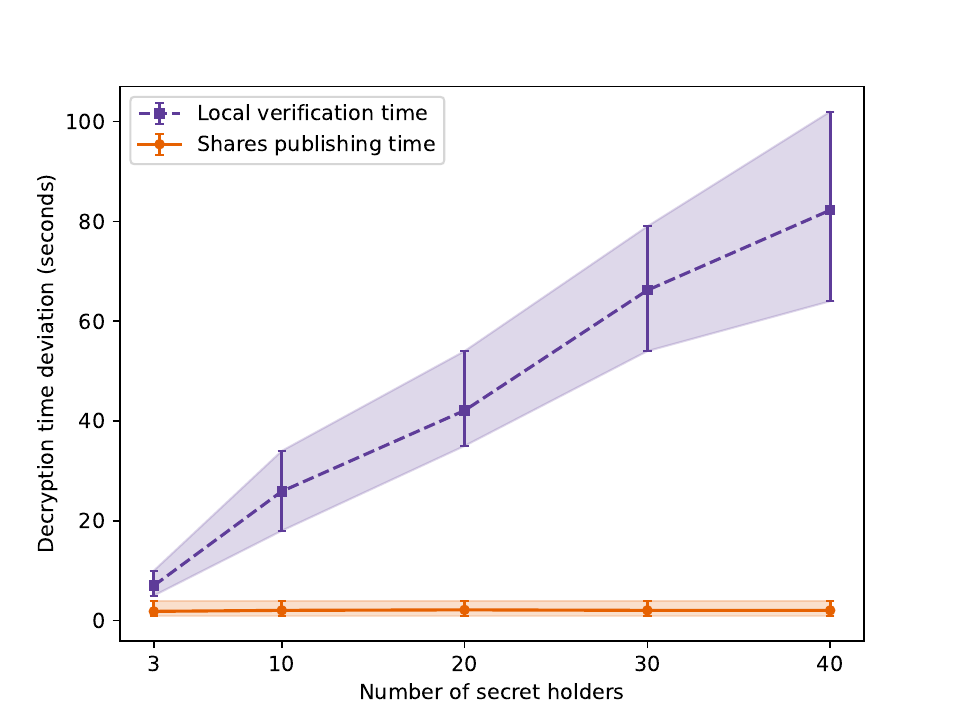}
\caption{Secret shares submission and verification latency at 3, 10, 20, 30, 40 secret holders}
\label{fig:scalability}
\end{figure}

In Figure \ref{fig:scalability}, the publishing time of the secret shares shows the latency of publishing a sufficient number of shares on the blockchain; the local verification time shows the latency of a secret holder getting enough secret shares verified. As holders process requests concurrently, increasing the number of holders does not extend the time required for the blockchain to accumulate a sufficient number of secret shares. However, as it requires more secret shares to validate a message when the number of secret holders scales, the time required for the secret holders to locally ensure the correct reconstruction of messages increases linearly.

In practice, the increase of local verification time for a secret holder can be improved by employing additional computation resources to verify secret shares instead of fully relying on the secret holders, given that the secret shares are publicly verifiable. Having a low latency in shares reveal means parties interested in certain messages can recover them with low latency, with an option to verify the correctness by themselves if they run faster computation than the secret holders.

\subsection{Energy Efficiency}
The testbed of the proposed system requires each secret holder to run the script~\cite{source_code} that occupies less than 10\% load on the server, which is estimated to consume 0.005 kWh~\cite{aws_power_consumption} given the server specifications and resource utilization rates. Such a system with 100 holders consumes 0.5 kWh. The power consumption of the system in a hundred years is less than the power consumption of Bitcoin~\cite{Bitcoin_power_consumption} (representing a time lock puzzle system) in one second.

\section{Resilience to Malicious Voting}
In this section, we demonstrate how the proposed timed-release cryptography preserves the integrity of an election in the presence of malicious voting attacks.

Cryptographic methods are employed to preserve the security of e-voting systems as they are susceptible to hacking activity and cyberattacks. These include safeguarding the privacy of voters~\cite{Asikis2020}, ensuring fairness in outcomes, allowing individuals to verify their votes, and enhancing the overall resilience of the system~\cite{zk-voting, smart_contract_voting, polys,proof_of_witness_presence}. 

Malicious voters can gain access to existing votes and can strategically vote for an alternative that is not ranked highest in their preference ordering to prevent a candidate from winning. This jeopardizes the election outcome and compromises the overall fairness of an election. Through established cryptographic methods such as bit commitments and blind signatures~\cite{blind_signature_voting}, voters maintain the confidentiality of their votes until the voting phase concludes. However, a drawback is that vote reveal is not automated and necessitates action from the holders. While some suggest combining zero-knowledge proofs and homomorphic encryption to keep ballots encrypted during voting~\cite{zk-voting, smart_contract_voting,polys}, it is not scalable due to the significantly high computation cost of homomorphic encryption. 

Therefore, time-based conditional information reveal is the only secure, automated, and scalable ballot protection method for electronic voting. Time-based conditional information reveal systems encrypt ballots till the voting phase ends, after which they are automatically decrypted by the network allowing relevant parties to tally them and determine the outcome.

With the idea of timed-release ballots, a system with accurate decryption time (especially resistant to early decryption) is vital. Otherwise, malicious voters can gain access to the early-decrypted ballots of a large population and jeopardize the election outcomes. In the following text, we model the malicious voting scenario considering real-world datasets and discuss if prevention is possible using the proposed system.

\paragraph{Simulating malicious voting}.
Each of the $\mathcal{N}$ voters selects from $\mathcal{K}$ alternatives. Each voter $i = 1,…, \mathcal{N}$ can provide a complete or partially complete strict preferential order over the alternatives. The plurality rule determines the winner.

{\em The sincere and malicious population of voters}: In every simulation, a certain population of voters turn into malicious and change their votes based on the winners of the remaining population. So if $l$\% of $\mathcal{N}$ voters are sincere, $(1-l)$\% of $\mathcal{N}$ voters change their votes to jeopardize the decision outcomes of the $l$\% of the voters. We have a total of 100 simulations, where $l$ is varied from 1 to 100, with an increment of 1. In every simulation, we have 100 iterations, wherein the set of malicious voters ($(100-l)$\%) changes due to random sampling. 

{\em Strategic voting by malicious voters}: Consider an election with $\mathcal{K}$=5 and $\mathcal{N}$ =100 voters. At each simulation, l is set to 80\%, so we have 20\% of malicious voters randomly sampled. Considering ($k_1$,$k_2$,$k_3$,$k_4$,$k_5$) alternatives, the aggregated preference of 80 voters (80\% of 100 voters) is $<$$k_3$,$k_4$,$k_5$, $k_2$,$k_1$$>$ and  $k_3$ is the winning alternative. A malicious voter had a true preference of $<$$k_5$,$k_1$,$k_3$,$k_2$,$k_4$$>$. However, the malicious voters have the intent to promote their first choice, while they demote and remove any candidate from their votes with higher aggregate preferences in the population of 80 sincere voters. This is also possible because the voters do not need to maintain a strict preferential order over all the alternatives. Following this, the malicious voter strategically changes their preference to $<$$k_5$,$k_1$,$k_2$$>$. The voter removes $k_3$,$k_4$ from the list as these candidates have a higher preference than the first choice ($k_5$) of the voter in the aggregated preference ($<$$k_3$,$k_4$,$k_5$, $k_2$,$k_1$$>$) of 80 voters.

{\em Generating decision outcomes}: The aggregated preference including malicious voters is calculated to assess if it deviates from the aggregated preference based only on sincere voters for all 100 iterations of a simulation. 
The probability of changing the winner is determined by the frequency of iterations in which the decision outcome changes due to malicious voting.

\paragraph{Impact on real-world voting datasets}
We consider real-world datasets for electing government leaders across different cities or countries (such as UK labor elections, elections in Dublin, Ireland, Oakland and Minneapolis elections in the United States~\cite{labor_voting_dataset, oakland_voting_dataset, dublin_voting_dataset, minneapolis_voting_dataset}, where voters provide either complete or incomplete strict preferential order for the candidates. The election data are taken from the {\em Preflib} library~\cite{mattei2013preflib}, which hosts benchmarks for real-world voting and preference data.

\begin{figure}[h!]
\centering
\includegraphics[width=9cm]{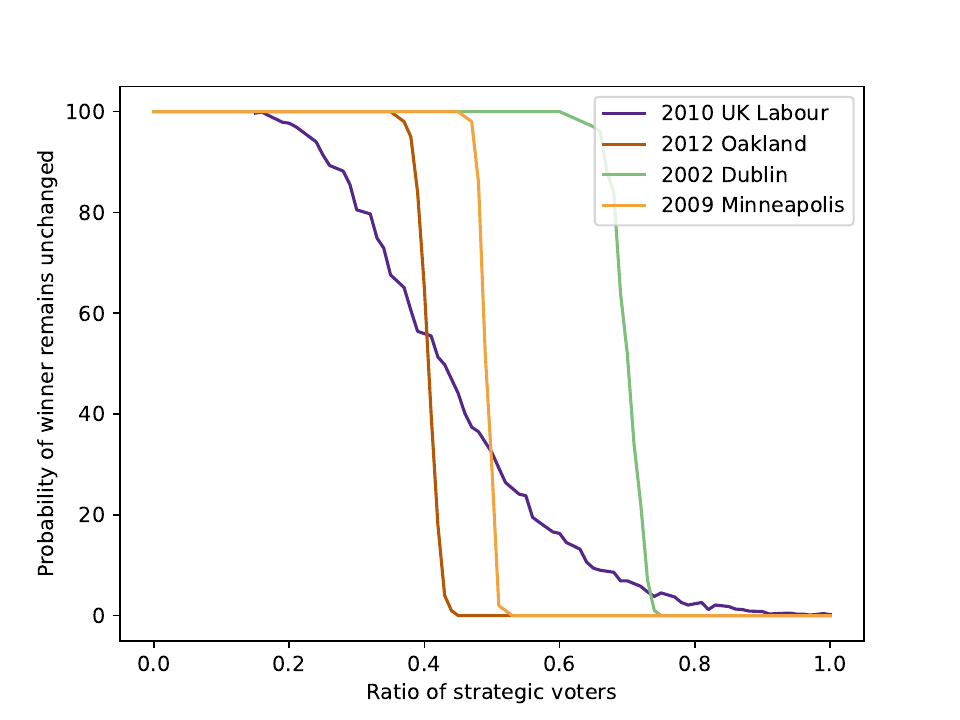}
\caption{Influence of voting result under different numbers of strategic voters}
\label{fig:strategic_voting}
\end{figure}

We select the datasets to have a mixture of elections that have different numbers of candidates (varying from 5 to 9) or voters (varying from 266 to 36655). As demonstrated in Figure \ref{fig:strategic_voting}, all four elections suffer from a change of winner in the presence of malicious voters. For example, the UK Labour Leadership Election, a small-scale election with only 266 voters, shows a change of winner in the present of 40 strategic voters. While larger-scale elections tolerate more malicious voters, a change of winner is still possible with the collaboration of potential malicious voters with similar preferences.

\paragraph{Decentralized timed-release ballot system to prevent malicious voting}
With a high probability of early decryption as shown in the previous section, time lock puzzle systems are prone to strategic voting. Conversely, the probability of early decryption in our proposed system is negligible. 

In the proposed system, for malicious voters to access existing ballots prematurely, they must acquire a sufficient number of secret shares before the preset decryption time. This would necessitate manipulating the blockchain clock so that the smart contract accepts the submitted secret shares before the end of voting. If this occurs, rational secret holders might submit their shares before voting ends, thereby enabling malicious voters to observe existing votes and cast their votes strategically.

Advancing the blockchain's time to the end of the voting period requires a malicious voter to also be a validator, and specifically, to be fortuitously selected as the block proposer for the critical block just before voting ends. For instance, in the Ethereum network, as illustrated in Section \ref{sec: 15s}, a malicious voter could potentially claim the block timestamp is advanced by up to 15 seconds. Given the current Ethereum network comprises over one million validators (as of May 2024)~\cite{eth_validators}, the likelihood of a malicious voter being chosen as the proposer for that precise block is negligible, less than 0.0001\%. Furthermore, even in the unlikely event that such an attack is initiated, the majority of honest secret holders still rely on their local, accurate clocks to submit their shares at the designated time, further complicating the feasibility of such an attack.

Moreover, the decentralized design of the proposed system allows sincere voters to become secret holders, collectively guarding each others' ballots until the end of voting. This further prevents malicious voters from compromising the system to gain early access to existing ballots.

In summary, a model is formulated to define malicious voting and apply it to real-world election datasets. This model is tested and validated on real-world datasets. We observe that if the proposed system is used, the probability of voters accessing or attacking the votes is negligible. Hence the proposed method is a secure solution to preserve fairness of real-world election outcomes.

\section{Conclusion and Future Work}
This paper illustrates a deep understanding of conditional information reveal systems and provides a resilient solution for secure timed-release cryptography in decentralized environments. Unlike existing proposals that mainly focus on the cryptographic protocols, our proposed timed-release cryptography system is not only secure against cryptographic attacks but also secure against various attacks against a timed-release cryptography system, including Sybil attack and time source poisoning. Moreover, we demonstrate the benefits of the proposed system through its application in e-voting scenarios, specifically examining how it mitigates the risks associated with the premature release of ballots in strategic voting contexts. The capability to prevent such scenarios shows the advantage of the proposed system in enhancing the integrity of e-voting processes. The relevance of our work extends beyond the specific domains of e-voting; it prompts further investigation into how the system could be integrated with various applications that rely on the dissemination of time-sensitive messages.

\subsection{Future Work}
While the proposed system embraces a high degree of decentralization, underpinned by blockchain technology and the involvement of multiple secret holders, a notable element of centralization persists. This centralized aspect involves reliance on a global clock as the primary time reference. A promising avenue for future research lies in a fully decentralized timekeeping mechanism characterized by elevated precision and reliability, which further improves the security of the system.

The proposed system operates as a timed-release messaging service, enabling the public broadcast of messages upon release. However, it does not specifically explore the implementations and applications of receiver-specified timed-release messages. Therefore, another potential direction is to expand the applicability of the system to offer timed-release messaging services that are only accessible to certain receivers, restricting the message availability to specific parties following the release of messages. 

\section{Acknowledgement}
This project is funded by a UKRI Future Leaders Fellowship (MR-/W009560-/1): `\emph{Digitally Assisted Collective Governance of Smart City Commons--ARTIO}'.

\bibliography{references}
\bibliographystyle{ieeetran}

\newpage
\appendix

\subsection{Numerical Example for the Cryptographic Protocol}
An example of the cryptographic protocol with small numbers is provided below for a better understanding of the process. The verification process using bilinear pairing and $G_2$ is excluded as it is not applicable in finite fields over small prime numbers. 

Suppose $G_1$ is a multiplicative group of integers modulo 23 and $g_1 = 11$. Consider there are 4 secret holders with private keys: $sk_1 = 3, sk_2 = 4, sk_3 = 5, sk_4 = 6$. Their public keys $pk_n = g_1^{sk_n}$ are therefore $pk_1 = 20$, $pk_2 = 13$, $pk_3 = 5$, $pk_4 = 9$. 

A client has a secret $k = 22$ to send as a timed-release message. The client generates $r = 7$ and computes $s_1 = pk_1^r = 21$, similarly $s_2 = 9, s_3 = 17, s_4 = 4$. The client uses Lagrange interpolation to derive a polynomial $P$ using three points $(0, 22), (1, 21),(2, 9)$, the resulting polynomial $P$ is $y = 6x^2 + 16x + 22$. The client evaluates $P(3) = 9, P(4) = 21$, and encrypts these two points using $s_3, s_4$ as keys, XOR as the encryption function to get ciphertexts $\alpha_3, \alpha_4$. $P(3)$ in binary is 1001, $s_3$ in binary is 10001, resulting in $\alpha_3 = 11000_{(2)} = 24_{(10)}$. Similarly, $\alpha_4 = 10001_{(2)} = 17_{(10)}$. The client can then share $g_1^r = 7$, $\alpha_3 = 24$, $\alpha_4 = 17$. 

Secret holders calculate their secret shares with $g_1^{r\cdot sk}$, and publish their secrets after the specified decryption time. The secret $k = 22$ can now be reconstructed with any three secret shares. For example, given $s_1, s_2, s_3$, we first calculate $P(3) = s_3 \oplus \alpha_3 = 17 \oplus 24 = 9$, then reconstruct the polynomial $P$ by interpolating $(1, 21), (2, 9), (3, 9)$, resulting in the same polynomial $P = 6x^2 + 16x + 22$, and evaluate at $x=0$ to get the number 22.

\end{document}